\documentclass[10pt]{iopart}
\usepackage{epsfig}
\usepackage{cite}

\begin{document}

\title{Superconductivity in tantalum self-intercalated 4$Ha$-Ta$_{1.03}$Se$_2$}

\author{Hua Bai$^1$, Mengmeng Wang$^1$, Xiaohui Yang$^1$, Yupeng Li$^1$, Jiang Ma$^1$, Xikang Sun$^1$, Qian Tao$^1$, Linjun Li$^2$, Zhu-An Xu$^{1,}$$^{3,}$$^{4,}$$^{5}$\footnote[3]{Corresponding
author. Tel: (86)571-87953255 E-mail address: zhuan@zju.edu.cn (Z.A.Xu)}}

\address{$^1$Department of Physics, Zhejiang University, Hangzhou 310027, P. R. China}

\address{$^2$State Key Laboratory of Modern Optical Instrumentation, College of Optical Science and Engineering, Zhejiang University, Hangzhou 310027, China}

\address{$^3$State Key Lab of Silicon Materials, Zhejiang University, Hangzhou 310027, P. R. China}

\address{$^4$Zhejiang California International NanoSystems Institute, Zhejiang University, Hangzhou 310058, P. R. China}

\address{$^5$Collaborative Innovation Centre of Advanced Microstructures, Nanjing 210093, P. R. China}

\ead{zhuan@zju.edu.cn}
\date{\today}

\begin{abstract}
TaSe$_2$ has several different polytypes and abundant physical
properties such as superconductivity and charge density waves
(CDW), which had been investigated in the past few decades.
However, there is no report on the physical properties of 4$Ha$
polytype up to now. Here we report the crystal growth and
discovery of superconductivity in the tantalum self-intercalated
4$Ha$-Ta$_{1.03}$Se$_2$ single crystal with a superconducting
transition onset  temperature $T_{\rm c}$ $\approx$ 2.7 K, which
is the first observation of superconductivity in 4$Ha$ polytype of
TaSe$_2$. A slightly suppressed CDW transition is found around 106
K. A large $\mu_0H_{\rm c2}/T_{\rm c}$ value of about 4.48 is
found when magnetic field is applied in the $ab$ plane, which
probably results from the enhanced spin-orbit coupling(SOC).
Special stacking faults are observed, which further enhance the
anisotropy. Although the density of states at the Fermi level is
lower than that of other polytypes, $T_{\rm c}$ remains the same,
indicating the stack mode of 4$Ha$ polytype may be beneficial to
superconductivity in TaSe$_2$.

\end{abstract}

\maketitle

\setlength{\parskip}{0\baselineskip}

\ioptwocol

\section{Introduction}
Transition metal dichalcogenides (TMDCs) are the $TX_2$-type
compounds ($T$ = transition metal, $X$ = S, Se, Te), which exhibit
many fascinating properties, such as
superconductivity\cite{morris1972superconductivity}, CDW
\cite{SUZUKI19841039}, large
magnetoresistance\cite{ali2014large}, and topological
semimetals\cite{deng2016experimental}, etc. TMDCs have a layered
structure, in which each layer consists of a $X-T-X$ sandwich. The sandwich
have two different types: octahedral and trigonal. The layers are
coupled  by van der Waals forces. Due to different sandwich types
and different stack modes, TMDCs often have many
polytypes\cite{katzke2004phase}, such as 1$T$, 2$Ha$, 2$Hb$,
3$R$, etc. The number dictates the layers number in one unit cell,
and the capital letter represents crystal system, i.e., $T$ is
tetragonal, $H$ is hexagonal, and $R$ is rhombohedral. Hexagonal TMDCs
also have different types, therefor use the lowercase letter to
distinguish them. In TMDCs, the modulation of CDW state always
lead to interesting phenomenon. For instance, with Cu
intercalation or electric-field gating effect, the CDW transition
of 1$T$-TiSe$_2$ can be suppressed and superconductivity
appears\cite{E2006Superconductivity,Li2016Controlling}. By
gate-controlled Li ion intercalation, 1$T$-TaS$_2$ thin flakes
undergo multiple phase transitions including the transition from
the Mott insulator to superconductor\cite{Yijun2015Gate}. In
2$Ha$-TaS$_2$, pressure suppresses CDW transition and
superconducting $T_{\rm c}$ is significantly
enhanced\cite{freitas2016strong}.

1$T$-TaSe$_2$ undergoes an incommensurate-charge-density waves
(ICCDW) phase transition at $T_{\rm ICDW}$ $\approx$ 473 K, and no
superconducting transition is
observed\cite{doi:10.1080/14786437408207261}. With S or Te doping,
superconductivity
appeares\cite{liu2016nature,liu2013superconductivity,ang2013superconductivity}.
2$Ha$-TaSe$_2$ also undergoes an ICCDW phase transition at $T_{\rm
ICDW}$ $\approx$ 122 K, then locks-in to a
commensurate-charge-density waves (CCDW) state at $T_{\rm CCDW}$
$\approx$ 90 K\cite{PhysRevB.16.801,PhysRevLett.45.576}, and
finally to a superconducting state at $T_{\rm c}$ $\approx$ 0.14
K\cite{Kumakura1996,YOKOTA2000551}. $T_{\rm c}$ will increase with
S or Te doping\cite{li2017superconducting,luo2015polytypism}, or
Ni or Pd intercalations \cite{Li20102248,bhoi2016interplay}, or
under pressure\cite{freitas2016strong}. 3$R$-TaSe$_2$ is not
stable\cite{bjerkelund1967structural}, but with Te, Mo or W
doping, it becomes stable and superconducting at $T_C$ of about 2
K \cite{0953-8984-27-36-365701}. 4$Hb$-TaSe$_2$ also has two
different CDW phase transitions at 410 K and 75 K respectively
\cite{PhysRevB.14.1543}, and 6$R$-TaSe$_2$ has the similar
behavior\cite{FUNG198047}. In 4$Hb$-TaS$_{2-x}$Se$_x$(0 $\leq$ x
$\leq$ 1.5) and 6$R$-TaS$_{2-x}$Se$_x$(0 $\leq$ x $\leq$ 1.6),
superconductivity and CDW coexist in whole composition range
\cite{doi:10.1063/1.4863311,doi:10.1063/1.4919219}. Nevertheless
for 4$Ha$-TaSe$_2$, there is few reports on its physical
properties up to now, most likely because this polytype is
difficult to synthesis.

Here we report the successful crystal growth and superconductivity
in Ta self-intercalated 4$Ha$-Ta$_{1.03}$Se$_2$ single crystal.
$T_{\rm c}$ of 4$Ha$-Ta$_{1.03}$Se$_2$ is about 2.7 K, which is
approximately 20 times higher than that of 2$H$-TaSe$_2$, close to
the $T_{\rm c}$ value reported for other doped or intercalated
polytypes. We also found a phase transition around 106 K, which is
analogous to a slightly suppressed CDW transition. A large
$\mu_0H_{\rm c2}/T_{\rm c}$ value of about 4.48 is found when
magnetic field is applied in the $ab$ plane which may be related
to the enhanced spin-orbit coupling (SOC).

\section{Experiment Details}

Single crystals of 4$Ha$-Ta$_{1.03}$Se$_2$ were made via  vapor transport technique using iodine as the transport agent. First, Ta(99.99\%) and Se(99.999\%) powder were mixed with the ratio of 1.06:2 and ground adequately, sealed into an evacuated quartz ampoule, heated to 800 $^{\circ}$C and kept for 3 days. Subsequently, the mixture was reground and added with iodine in a concentration of 8.6 mg/cm$^3$, sealed into an evacuated quartz ampoule with a length of 16 cm. The ampoule was heated for 7 days in a two-zone furnace, where the temperature of source zone and growth zone were fixed at 850 $^{\circ}$C and 800 $^{\circ}$C respectively. Finally, the silver, mirror-like plates single crystals with typical size of about 4 $\times$ 4 $\times$ 0.05 mm$^3$ were obtained.
We also tried to fabricate the stoichiometric 4$Ha$-TaSe$_2$ with the same method but failed. We guess that the redundant Ta atoms conduce to the formation of $4Ha$ polytype.

The single crystal X-ray diffraction (XRD) data were collected by
a PANalytical X-ray diffractometer (Empyrean) with a Cu K$_\alpha$
radiation and a graphite monochromator at room temperature. The
chemical compositions were determined by energy-dispersive X-ray
spectroscopy (EDX) with a GENESIS4000 EDAX spectrometer.
High-resolution transmission electron microscope (HRTEM) images
were taken at room temperature with an aberration corrected
FEI-Titan G2 80-200 ChemiSTEM. DC magnetization was measured on a
magnetic property measurement system (MPMS-XL5, Quantum Design),
and the ``Ultra Low Field'' option was used. The specific-heat
capacity was measured on a physical properties measurement system
(PPMS-9, Quantum Design), using a relaxation technique. The
electrical properties were measured on an Oxford Instruments-15T
cryostat with a He-3 probe.

\section{Result and Discussion}

\begin{figure}[!thb]
\begin{center}
\includegraphics[width=3in]{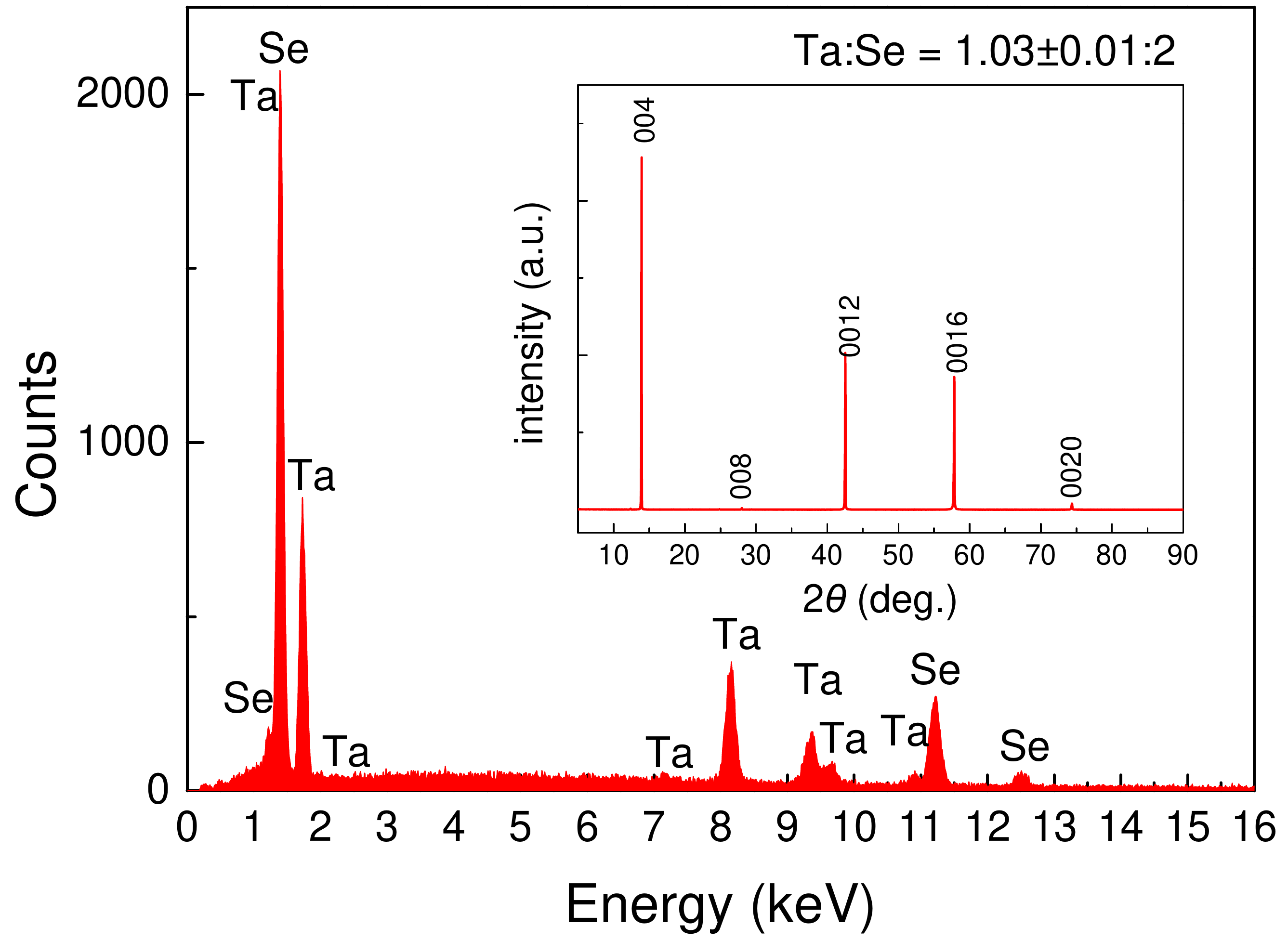}
\end{center}
\caption{\label{Fig1}EDX pattern of 4$Ha$-Ta$_{1.03}$Se$_2$. Inset: XRD pattern of a 4$Ha$-Ta$_{1.03}$Se$_2$ single crystal.}
\end{figure}

Figure \ref{Fig1} shows the EDX pattern of the single crystal. The
ratio of Ta to Se is determined to be 1.03(1):2, which is
an average result for multiple areas in the same single crystal.
Although the starting ratio is 1.06:2, the redundant amount of Ta
atom are not all intercalated into the sample. We get similar
results for several batches under the same condition. The room
temperature XRD pattern of a 4$Ha$-Ta$_{1.03}$Se$_2$ single
crystal is shown in the inset of Figure \ref{Fig1}, where all the
reflections are (00$l$) peaks. The lattice constant along the
$c$-axis is calculated to be 25.436 \AA. Comparing with 25.180
\AA \  of 4$Ha$-TaSe$_2$ from the literature\cite{Bjerkelund},
the c-axis is a little expanded. This should result from the Ta
self-intercalation, similar to the Ni or Pd intercalated
2$H$-TaSe$_2$\cite{Li20102248,bhoi2016interplay}.

\begin{figure*}[!thb]
\begin{center}
\includegraphics[width=7in]{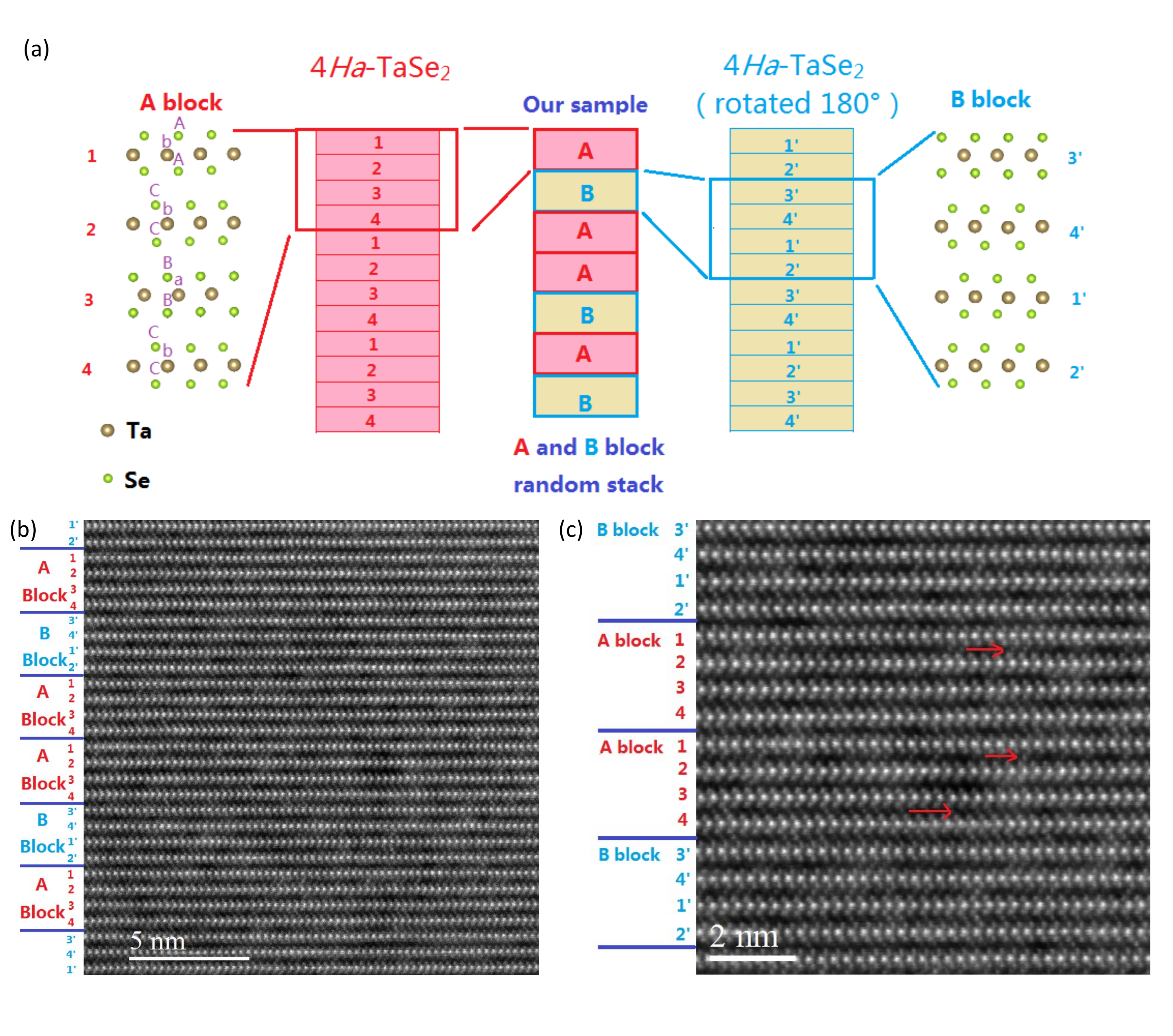}
\end{center}
\caption{\label{Fig2}(a) The structure of 4Ha polytype and the
illustration for the special stacking faults in our samples.
The purple letters in the left structure illustrate
the stacking sequence of 4$Ha$ polytype.
The right structure of 4$Ha$-TaSe$_2$ is rotated $180^{\circ}$ along
the $c$-axis. (b) HRTEM image of a 4$Ha$-Ta$_{1.03}$Se$_2$ single
crystal from zone axes [100]. (c) Locally enlarged HRTEM image.
The red arrows point out some of the intercalated Ta atoms.}
\end{figure*}

The structure model is displayed in Figure \ref{Fig2}(a), and
Figure \ref{Fig2}(b) and (c) shows the HRTEM images of a single
crystal from zone axes [100]. The HRTEM images clearly reveal the
atoms arrangement mode and thus provide a direct evidence for the
4$Ha$ polytype. Every 4 layers constitute one unit cell. The layer
stacking type can be described by using the method from the
literature\cite{katzke2004phase}, where A(a), B(b), C(c) represent
three positions of atoms in the $ab$-plane. The capital letters
correspond to the Se atoms and the lower case letters designate
the Ta atoms. The same letters, no matter capital or lower case,
represent the same positions in the $ab$-plane for all layers.
Every three letters describe one sandwich layer. Along the
$c$-axis, the stacking sequence of 4$Ha$ polytype is AbACbCBaBCbC,
just as illustrated in the left structure in Figure \ref{Fig2}(a).
The HRTEM images also reveal special stacking faults in our
samples as displayed in Figure \ref{Fig2}(a). Both A block and B
block consist of 4 layers of the 4$Ha$-TaSe$_2$ structure, but
from different sections and directions as shown in Figure 2(a).
Our sample is randomly stacked by the A and B blocks. These
stacking faults could enhance the anisotropy due to the
2-dimensional characteristic of fragments. Moreover, the
intercalated Ta atoms between two layers can be clearly observed
in the locally enlarged HRTEM image as shown in Figure
\ref{Fig2}(c). It can be found that the Ta intercalation is
microscopically not homogeneous.

\begin{figure*}[!thb]
\begin{center}
\includegraphics[width=6in]{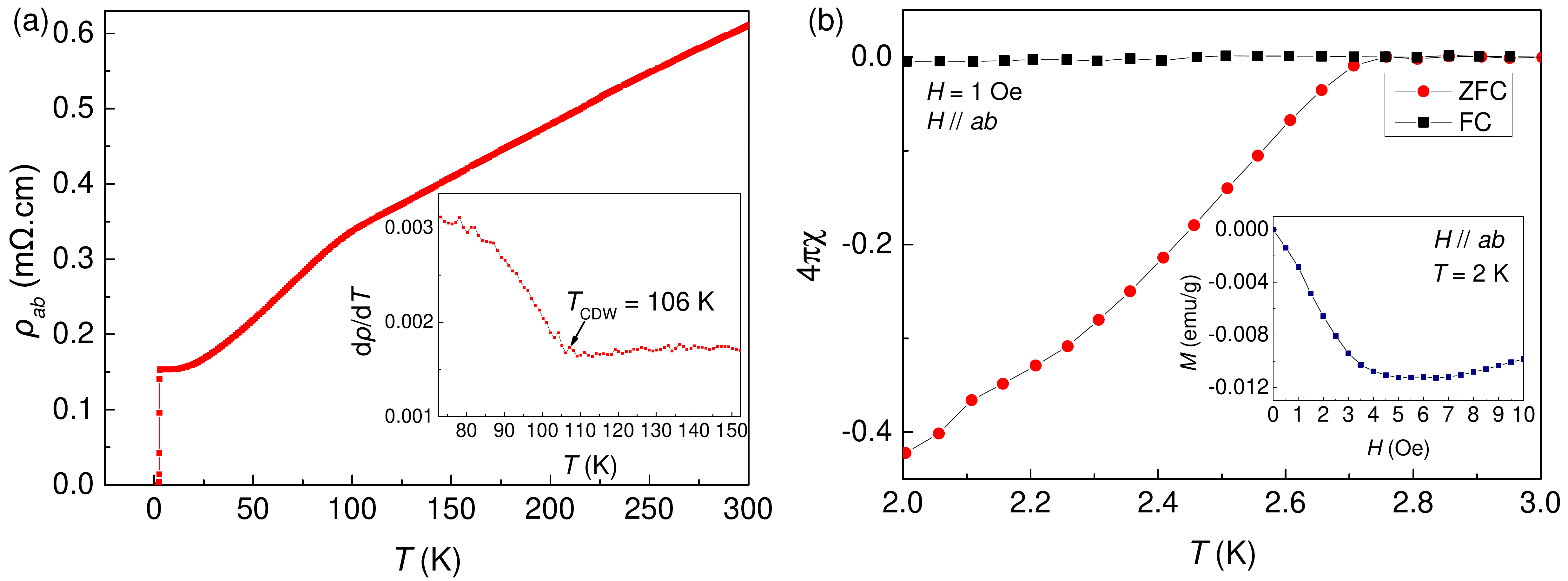}
\end{center}
\caption{\label{Fig3}(a) Temperature dependent in-plane
resistivity of 4$Ha$-Ta$_{1.03}$Se$_2$ for 1.5 - 300 K. Inset:
Derivative of resistivity of 4$Ha$-Ta$_{1.03}$Se$_2$.
(b) Temperature dependence of dc magnetic
susceptibility for 4$Ha$-Ta$_{1.03}$Se$_2$ ($H//ab$, $H$ = 1 Oe). ZFC
and FC denote zero-field cooling and field cooling. Inset: The field dependence of initial part of magnetization curve ($H//ab$, $T$ = 2 K).}
\end{figure*}

Figure \ref{Fig3}(a) displays the temperature dependence of
in-plane electrical resistivity for 4$Ha$-Ta$_{1.03}$Se$_2$ single
crystals. The sample was cut to a size of about 4 $\times$ 1 $\times$ 0.05 mm$^3$
for the resistivity measurement. Similar to other polytypes, 4$Ha$-Ta$_{1.03}$Se$_2$ are
metallic in normal state. Around 106 K, a slope change appears.
This change can be better observed in the derivative of resistivity curve
 as shown in the inset. Below 106 K, there is a
suddenly increase in the derivative of resistivity curve. This
slope change is analogous to the case in other TMDCs and usually
it is attributed to the CDW transitions\cite{bhoi2016interplay}.
The onset superconducting transition temperature $T_{\rm c}$  is
about 2.7 K, with a narrow transition width of about 0.2 K. For
other polytypes, when doping or intercalating, $T_{\rm c}$ is
always in the range of 2 - 4 K. Although 4$Ha$ polytype has a
disparate stacking pattern, the superconducting transition
temperature remains similar to the others. Figure \ref{Fig3}(b)
shows the results of magnetic measurements. The inset is the field
dependence of initial part of magnetization curve at 2 K with
magnetic field applied parallel to the long direction in the
$ab$-plane. The lower critical field ($H_{c1}$) is estimated to be
about 3 Oe. Thus we measured the temperature dependence of dc
magnetic susceptibility under magnetic field of 1 Oe applied in
the $ab$-plane. It was measured for both zero-field cooling (ZFC)
and field cooling (FC). The $T_{\rm c}$ determined from magnetic
susceptibility is 2.7 K, close to the onset temperature derived
from the resistivity transition. The estimated superconducting
shielding volume fraction is about 42.2\% at 2 K. As mentioned
above, the Ta intercalation is microscopically not homogeneous,
therefore we conjecture that intercalation concentration is
important for the occurrence of superconductivity,
 i.e., only the area with enough intercalated Ta atoms will show superconductivity.

\begin{figure*}[!thb]
\begin{center}
\includegraphics[width=6in]{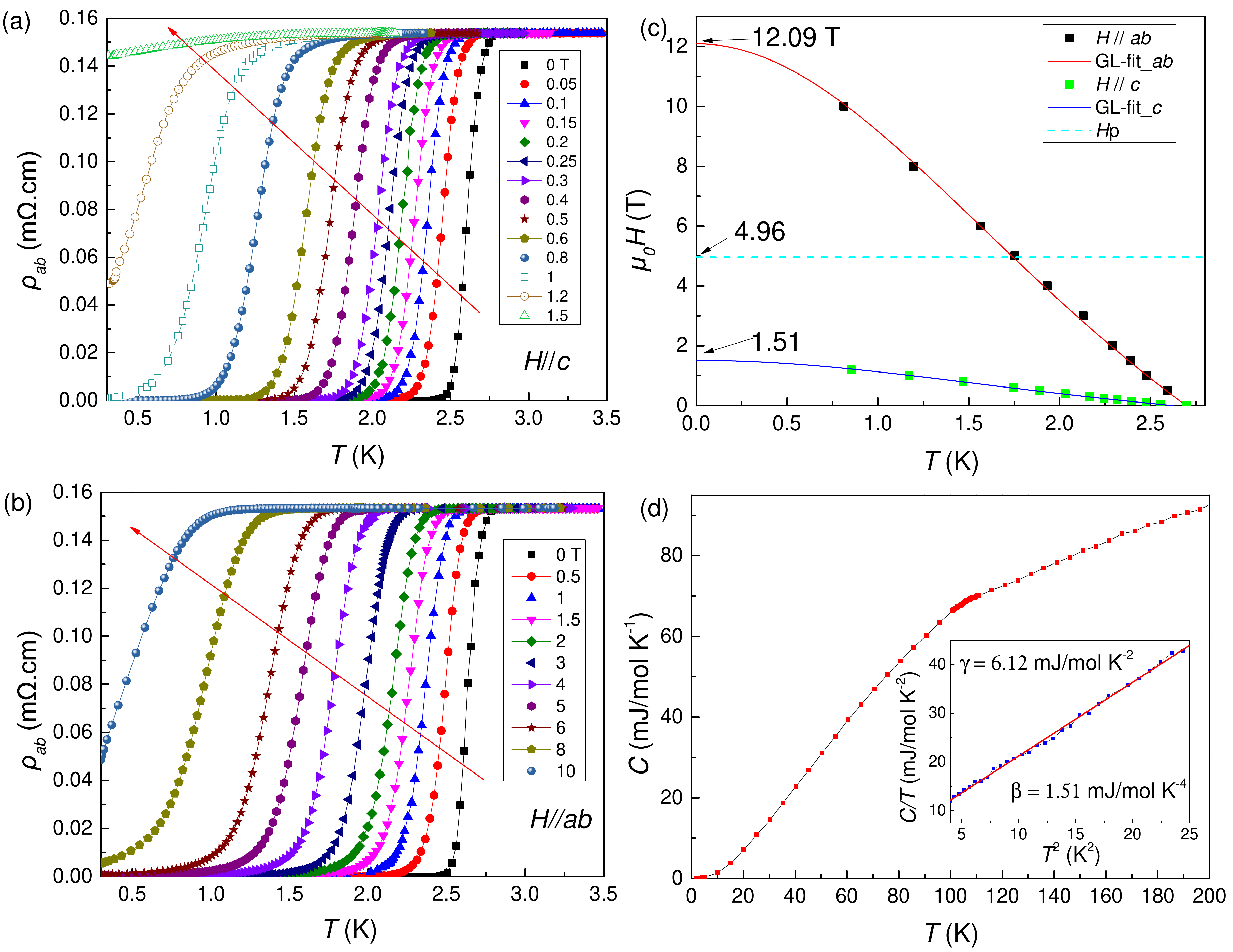}
\caption{\label{Fig4}(a) (b) The resistivity under different
magnetic fields from 0.3 - 3.5 K for 4$Ha$-Ta$_{1.03}$Se$_2$.
(a)$H//c$, up to 1.5 T. (b)$H//ab$, up to 10 T. (c) The upper
critical field and the G-L fits for both field directions. The dash
 cyan line dictates the Pauli paramagnetic limit.(d)
Temperature dependence of specific heat from 1.8 K - 200 K for
4$Ha$-Ta$_{1.03}$Se$_2$. Inset: Low temperature part of specific
heat divided by temperature, $(C/T)$, as a function of $T^2$, and
the red line is a linear fit.}
\end{center}
\end{figure*}

Figure \ref{Fig4}(a) and (b) display the temperature dependence of
in-plane electrical resistivity under different magnetic fields of
4$Ha$-Ta$_{1.03}$Se$_2$ with field applied parallel to the
$c$-axis and in the $ab$-plane respectively. For both field
directions, the resistivity curves shift parallel down towards
the low temperature with the increasing fields. This behavior
is different from intercalated 2$H$-TaSe$_2$ whose resistive
transition broadens when magnetic field is applied out of the
 $ab$-plane\cite{bhoi2016interplay}. We
summarize the temperature dependent upper critical
field ($H_{c2}$) determined by $T_{\rm c}^{\rm onset}$ (90\%
of normal state resistivity) in Figure \ref{Fig4}(c). In both
directions, the $H_{c2}$ data can be well fit by the
Ginzberg-Landau (GL) equation\cite{PhysRev.147.295}:
\begin{equation} %\begin{split}
\label{1}
%\begin{align}
H_{c2}(T)=H_{c2}(0)(\frac{1-t^2}{1+t^2}) %\label{1}
%\end{align}
%\end{split}
\end{equation}
where $t=T/T_{\rm c}$ is the reduced temperature. From this fit we
can deduce  $\mu$$_0H_{c2}^c$(0) = 1.51 T and
$\mu$$_0H_{c2}^{ab}$(0) = 12.09 T. The Pauli paramagnetic limit
for the upper critical field is $\mu$$_0H_P$ = 1.84$T_{\rm c}$ =
4.96 T. $H_{c2}^{ab}$(0) is about 2.44$H_P$, where
$\mu$$_0H_{c2}^{ab}$(0)/$T_{\rm c}$ = 4.48. For 2$H$-TaSe$_2$,
$H_{c2}$ is much smaller than Pauli paramagnetic limit for
both directions, and $\mu$$_0H_{c2}^{ab}$(0)/$T_{\rm c}$ = 0.03
only\cite{YOKOTA2000551}. However, for the Pd intercalated
2$H$-Pd$_{0.09}$TaSe$_2$, $\mu$$_0H_{c2}^{ab}$(0)/$T_{\rm c}$ =
3.75\cite{bhoi2016interplay}, just slightly lower than the value
in this work. Such a large value of $\mu$$_0H_{c2}$(0)/$T_{\rm c}$
 was also found in quasi-one-dimensional Nb$_2$PdS$_5$ where the value is
3\cite{Q2013Superconductivity}, and in Pt, Ir or Ru doped Nb$_2$PdS$_5$
where $\mu$$_0H_{c2}$(0)/$T_{\rm c}$ increases  due to
additional SOC effect\cite{Zhou2014,Chen2017Enhanced}. We speculate that
in TaSe$_2$, extra Ta atoms may resemble the effect to increase
SOC, hence lead to the large upper critical fields. The anisotropy
factor $\gamma$ =  $H_{c2}^{ab}(0)$/$H_{c2}^c(0)$
$\approx$ 8, which is larger than the value of about 3 - 4 for 2$H$
polytypes\cite{YOKOTA2000551,bhoi2016interplay}. As
mentioned above, the special stacking faults may also contribute
to such large anisotropy.

Figure \ref{Fig4}(d) shows the temperature dependence of specific-heat $C(T)$ for 4$Ha$-Ta$_{1.03}$Se$_2$ from 1.8 K to 200 K.
 No distinct specific-heat jump is observed for the CDW transition around 106 K. Combining with the minimum in derivative
 curvature of resistivity, we can conclude that the CDW transition has been severely suppressed by Ta-self intercalation
  in 4$Ha$-Ta$_{1.03}$Se$_2$. This phenomenon is similar to 2$H$-Ni$_{0.02}$TaSe$_2$\cite{Li20102248} where there is an
  anomaly in temperature dependent resistivity but not in the specific-heat. Moreover, similar to 2$H$-Ni$_{0.02}$TaSe$_2$,
  there is no obvious specific-heat jump around $T_{\rm c}$, which may be due to the low volume fractions of magnetic shielding.
 The $C/T$ versus $T^2$ curve is displayed in the insert of Figure \ref{Fig4}(d).
The data can be well fit by the equation $C/T$ =
$\gamma$+${\beta}T^2$,  where $\gamma$ is Sommerfeld coefficient
for electronic specific-heat and $\beta$ is lattice specific-heat
coefficient. From the fit results we can obtain $\gamma$ = 6.12
mJ/mol K$^{-2}$ and $\beta$ = 1.51 mJ/mol K$^{-4}$. Furthermore,
we can estimate the Debye temperature $\Theta$$_D$ by  the formula
 $\Theta$$_D$=(12$\pi$$^4nR/5\beta$)$^{1/3}$ = 157.42 K, where $n$ is the number of atoms per formula unit ($n$ = 3.03),
 and $R$ is the gas constant. The electron-phonon coupling constant $\lambda$$_{e-ph}$
 can be estimated by the McMillan equation\cite{PhysRev.167.331}:
\begin{equation} %\begin{split}
\label{2}
%\begin{align}
\lambda_{e-ph}=\frac{\mu^*ln(\frac{\Theta_D}{1.45Tc})+1.04}{ln(\frac{\Theta_D}{1.45Tc})(1-0.62\mu^*)-1.04} %\label{1}
%\end{align}
%\end{split}
\end{equation}
where $\mu^*$ is Coulomb pseudopotential and is often set to 0.15 as the empirical value.
From the estimation we get $\lambda$$_{e-ph}$ = 0.69. This value is less than the
minimum value 1 of strong coupling, which suggest it is still in an intermediate coupling range.
The density of states at the Fermi level $N(E_F)$ can be calculated by the equation:
\begin{equation} %\begin{split}
\label{3}
%\begin{align}
N(E_F)=\frac{3\gamma}{\pi^2k_B^2(1+\lambda_{e-ph})} %\label{1}
%\end{align}
%\end{split}
\end{equation}
$N(E_F)$ = 1.53 state/eV f.u. for our sample. This value is close to that of undoped 2$H$-TaSe$_2$
which is 1.51 state/eV f.u., but lower than those of other doped or intercalated TaSe$_2$ which are
about 2 state/eV f.u.\cite{bhoi2016interplay}. In general, high $N(E_F)$ is in favor of superconductivity.
In spite of having low $N(E_F)$, 4$Ha$-Ta$_{1.03}$Se$_2$ also has comparable $T_{\rm c}$ to other doped or intercalated TaSe$_2$,
 indicating that the stacking mode of 4$Ha$ may be in favor of superconductivity than other polytypes.

\section{Conclusion}
In summary, we discover superconductivity with an onset $T_{\rm
c}$ of 2.7 K in the Ta self-intercalated 4$Ha$-Ta$_{1.03}$Se$_2$.
 A suppressed CDW transition is observed around 106 K. A kind of special stacking faults is revealed,
 which may increase the anisotropy. The $\mu$$_0H_{c2}^{ab}$(0)/$T_{\rm c}$ = 4.48 is larger than other polytypes,
 and the estimated $H_{c2}^{ab}$(0) is 2.44 times of the Pauli paramagnetic limit, which may result from enhanced SOC.
 $\lambda$$_{e-ph}$ = 0.69, indicating that it is still in the intermediate coupling range. On the other hand,
 $N(E_F)$ is only 1.53 state/eV f.u., smaller than that of other doped or intercalated TaSe$_2$,
 but $T_{\rm c}$ is almost the same as them. This work indicates that the stack mode of 4$Ha$ polytype is of benefit to
 superconductivity and $H_{\rm c2}$ for TaSe$_2$, and superconducitivty could be even enhanced if $N(E_F)$ can be increased.

\section{Acknowledgments}
We thank Yi Zhou and Guanghan Cao for helpful discussions. This
work was supported by the National Basic Research Program of China
(Grant Nos. 2014CB92103) and the National Key R\&D Projects of
China (Grant No. 2016FYA0300402), the National Science Foundation
of China (Grant Nos. U1332209 and 11774305), and the Fundamental
Research Funds for the Central Universities of China. We thank the
Center of Electron Microscopy of ZJU for the access to microscope
facilities used in this work. We also thank Lei Qiao for the help
on using the Ultra Low Field option in the magnetic measurements.

\section{References}
\providecommand{\newblock}{}

\end{document}